\documentstyle{article}
\begin{document}

\baselineskip=23pt

\begin{center}
{\bf {\Huge An Accelerated Expansion Model in the Absence of the
Cosmological Constant}}

\vspace{6mm}

{\bf Yi-Ping Qin$^{1,2,3}$ }
\end{center}

{\bf $^{1}$ Yunnan Observatory, Chinese Academy of Sciences, Kunming, Yunnan
650011, P. R. China; E-mail: ypqin@public.km.yn.cn }

{\bf $^{2}$ National Astronomical Observatories, Chinese Academy of Sciences 
}

{\bf $^{3}$ Chinese Academy of Science-Peking University joint Beijing
Astrophysical Center }

\vspace{3mm}

\begin{center}
{\bf {\Large Summary}}
\end{center}

\begin{quote}
{\bf Based on some observations, the apparent energy, associated with
gravity, of vacuums is defined, with that of normal vacuums to be zero and
that of the vacuums losing some energy to be negative. An important
application of the energy is its contribution to Einstein's equation. A
cosmological model, accounting for recent observations of the accelerated
expansion of the universe, in the absence of the cosmological constant, can
be well constructed. In a certain case, the expansion of the universe would
be decelerated at its early epoch and accelerated at its late epoch. The
curvature of the universe would depend on the ratio of matter energy to
total energy. The missing mass problem does no longer exist in this model.
Most negative apparent energy vacuums might be contained in voids, then the
spacetime of galaxy clusters or that of the solar system would not be
significantly affected by this kind of energy.}
\end{quote}

\vspace{2mm}

\begin{quote}
\end{quote}

\vspace{10mm}

Recent observations showed that the expansion of the universe is accelerated
rather than decelerated [1--3]. An economic approach to this phenomenon is
to adopt the cosmological constant which is often referred to the vacuum
energy produced by the phase transitions the universe undergoes as it cools.
However, there are some reasons against this scenario [4]. For example, the
amount of vacuum energy produced by all the phase transitions can be about $%
10^{120}$ times greater than the density of all the matter in the universe
[5]; in particular, the constant corresponds to a universal energy density
but its influence on the nearby spacetime has never been observed. These
facts suggest that vacuum energy acts as something like potential energy and
the apparent energy associated with gravity, of the vacuums in the nearby
spacetime might be zero.

Recently, the Casimir force was detected in laboratory [6]. It is reasonable
that a work done by the force would extract more or less energy from
vacuums. We call a vacuum not losing any energy a normal vacuum and that
losing some a deficit vacuum. According to the above comprehension, we
define the apparent energy associated with gravity, of normal vacuums to be
zero and that of deficit vacuums to be negative. The energy is assumed to
contribute to Einstein's equation the way matter energy does.

Now we consider a cosmological model of the Robertson-Walker metric
following Einstein's equation and the conservation equation of the
energy-momentum tensor. The difference is that, we take 
\begin{equation}
\rho =\rho _m+\rho _v
\end{equation}
and 
\begin{equation}
p=p_m+p_v
\end{equation}
with that $\rho _v$ can be negative, where $m$ denotes matter and $v$
represents vacuums.

For the Robertson-Walker metric, Einstein's equation gives [7] 
\begin{equation}
3\stackrel{\bullet \bullet }{R}=-4\pi G(\rho +3p)R,
\end{equation}
\begin{equation}
R\stackrel{\bullet \bullet }{R}+2\stackrel{\bullet }{R}^2+2k=4\pi G(\rho
-p)R^2,
\end{equation}
and the conservation equation of the energy-momentum tensor yields 
\begin{equation}
\stackrel{\bullet }{p}R^3=\frac d{dt}[R^3(\rho +p)],
\end{equation}
where $\stackrel{\bullet }{R}=dR/dt$. From (3) and (4) one can obtain 
\begin{equation}
\stackrel{\bullet }{R}^2+k=\frac{8\pi G}3\rho R^2
\end{equation}
and 
\begin{equation}
2R\stackrel{\bullet \bullet }{R}+\stackrel{\bullet }{R}^2+k=-8\pi GpR^2.
\end{equation}
Let us define 
\begin{equation}
H\equiv \frac{\stackrel{\bullet }{R}}R,
\end{equation}
\begin{equation}
q\equiv -\frac{R\stackrel{\bullet \bullet }{R}}{\stackrel{\bullet }{R}^2},
\end{equation}
\begin{equation}
\rho _c\equiv \frac{3H^2}{8\pi G},
\end{equation}
and 
\begin{equation}
\Omega \equiv \frac \rho {\rho _c}.
\end{equation}
Then equations (6) and (7) can be written as 
\begin{equation}
\frac k{R^2}=H^2(\Omega -1)
\end{equation}
and 
\begin{equation}
\frac k{R^2}=H^2(2q-1-\frac{3p}\rho \Omega ),
\end{equation}
respectively.

The observation of $q<0$ suggests that, at least at the present time, the
following condition must be satisfied (see equation (3)): 
\begin{equation}
\rho +3p<0.
\end{equation}

One means to meet condition (14) is to consider a universe containing both
deficit and normal vacuums (then on the average, $\rho _v<0$), and to assume
that deficit vacuums act as negative energy photons (then $p_v=\rho _v/3<0$%
). (According to quantum electrodynamics, normal vacuums are full of all
kinds of electromagnetic modes.)

At the late epoch of the universe, the pressure of matter particles is
negligible. Then $p=p_v$, $3p=3p_v=\rho _v$. Condition (14) leads to $\rho
<-\rho _v$, which allows $\rho >0$ (note $\rho _v<0$). Hence, it is possible
that a positive energy density (where $\rho _m>-\rho _v$) may lead to an
accelerated expansion of the universe at its late epoch so long as $\rho
<-\rho _v$ or $\rho _m<-2\rho _v$. For $\rho >0$, we find $-\rho _v<\rho
_m<-2\rho _v$. Equation (13) leads to $k/R^2=H^2(2q-1-\alpha \Omega )$,
where 
\begin{equation}
\alpha \equiv \frac{\rho _v}\rho .
\end{equation}
This together with (12) yield $\Omega =2q/(1+\alpha )$. As $\rho <-\rho _v$,
for $\rho >0$, we find $\alpha <-1$. For $q<0$, this indicates that $\Omega
>0$. The curvature of the universe depends on $\alpha $: when $\alpha
>-(1-2q)/\Omega $, then $k=-1$ and $0<\Omega <1$; when $\alpha
=-(1-2q)/\Omega $, then $k=0$ and $\Omega =1$; when $\alpha <-(1-2q)/\Omega $%
, then $k=+1$ and $\Omega >1$. Equation (15) shows, if $\alpha $ is a
constant, the sign of $\rho $ will remain unchanged. For a flat universe, $%
\alpha =-(1-2q)$ at the late epoch. When $\alpha $ is known (e.g.,
determined by the late epoch acceleration), $\rho $ and $\rho _v$ may be
known since $\rho _m$ is measurable.

At the early epoch of the universe, $p_m=\rho _m/3$. Then $p=\rho /3$ (as $%
p_v=\rho _v/3$). For $\rho >0$, we find from (3) that $\stackrel{\bullet
\bullet }{R}<0$ (then $q>0$), indicating that the expansion of the universe
is decelerated. Equation (13) leads to $k/R^2=H^2(2q-1-\Omega )$. This
together with (12) yield $\Omega =q$. Since $q>0$, then $\Omega >0$. When $%
k=-1$, then $2q-1<\Omega <1$, $0<q<1$; when $k=0$, then $\Omega =1$, $q=1$;
when $k=+1$, then $1<\Omega <2q-1$, $q>1$.

We find in this model that the universe can possess a positive energy
density ($\rho >0$, where $\rho _m>-\rho _v$) and a positive energy
parameter ($\Omega >0$). The curvature of the universe depends on the ratio
of vacuum energy (or matter energy) to total energy. As the pressure of
matter particles becomes less important as time goes on, the expansion of
the universe will change. When $p_m=((\alpha +1)/3(\alpha -1))\rho _m$, we
find from (3) that $\stackrel{\bullet \bullet }{R}=0$. It is at this moment
the expansion changing from deceleration to acceleration.

It is obvious that deficit vacuums, as they possess negative energy, will be
expelled by gravitation. Then there will seldom be deficit vacuums remained
in galaxy clusters, and therefore the spacetime of galaxy clusters and that
of the solar system will not be significantly affected. We suspect that the
place in the universe containing most deficit vacuums might be voids. In the
above model, deficit vacuums are assumed to act as negative energy photons.
These photons will be deflected towards the center of voids and then the
voids might be crowded with them. Within the voids, if the amount of vacuum
apparent energy (negative) is over that of matter energy, the spacetime
might be somewhat like that of the Schwarzschild solution with (more or
less) negative mass, and then matter objects would experience (strong or
weak) anti-gravitation. Many matter objects must have been driven to the
sheets around the voids, and the number of those remained will be small. In
addition, any matter objects would undergo the negative pressure ($%
p(void)\simeq p_v(void)=\rho _v(void)/3<0$ ) in the voids. Those survived
would be that with their components being firmly connected. The cloud
structure of matter must finally be disintegrated and some other matter
structures must be reduced. As a result, the amount of matter in the sheets
would grow and that in the voids would reduce (the disintegrated matter
would be easier to be expelled to the sheets by anti-gravitation). In this
model, we can assign the observed matter density to be $\rho _m$ (e.g., at
the present time, taking $\rho _m=0.3\rho _c$ [8]), then the missing mass
problem will no longer exist. (We suggest that many conventional problems
should be reexamined in this new model.)

It might be possible that deficit vacuums act as negative mass particles, or
some act as negative mass particles while others act as negative energy
photons. Situations in these cases will be different.

\vspace{30mm}

{\bf ACKNOWLEDGEMENTS}

The author is grateful to Professors G. Z. Xie, Xue-Tang Zheng and Shi-Min
Wu for their guide and help. This work was supported by the United
Laboratory of Optical Astronomy, CAS, the Natural Science Foundation of
China, and the Natural Science Foundation of Yunnan.

\newpage

\begin{verse}
{\bf REFERENCES}

1. Garnavich, P. M. et al. (1998). {\it Astrophys. J.} {\bf 493}, L53.

2. Perlmutter, S. et al. (1998). {\it Nature} {\bf 391}, 51.

3. Riess, A. G. et al. (1998). {\it Astron. J.} {\bf 116}, 1009.

4. Coles, P. (1998). {\it Nature} {\bf 393}, 741.

5. Weinberg, S. (1989). {\it Rev. Mod. Phys. }{\bf 6}, 1.

6. Lamoreaux, S. K. (1997). {\it Phys. Rev. Lett.} {\bf 78}, 5.

7. Weinberg, S. (1972). In {\it Gravitation and Cosmology}, (John Wiley, New
York).

8. Trimble, V., and Aschwanden, M. (1999). {\it Publ. Astron. Soc. Pac. }%
{\bf 111}, 385.
\end{verse}

\end{document}